# Superstrate structured $Sb_2S_3$ thin-film solar cells by magnetron sputtering of Sb and post-sulfurization

Evgeniia Gilshtein[a], Harshvardhan Maheshkant Gupta[a], Andrea Maria Pierri Enevoldsen[a], Cristina Besleaga[b], Aurelian Catalin Galca[b,c], Stela Canulescu[a]

[a] Department of Electrical and Photonics Engineering, Technical University of Denmark, Roskilde 4000, Denmark

[b] Laboratory of Complex Heterostructures and Multifunctional Materials (HeCoMat), National Institute of Materials Physics, Magurele 077125, Romania

[c] International Center for Advanced Training and Research in Physics, Magurele 077125, Romania

**Abstract**

This study explores the fabrication and optimization of superstrate-structured antimony sulfide ($Sb_2S_3$) thin-film solar cells using RF magnetron sputtering of antimony (Sb) followed by post-sulfurization. The study systematically investigates the effects of varying absorber and buffer layers thicknesses on the photovoltaic performance of FTO/CdS/$Sb_2S_3$/Spiro-OMeTAD/Au solar cell devices. Optimizing the $Sb_2S_3$ absorber thickness to 100 nm resulted in a maximum power conversion efficiency of the champion device of 2.76%, with enhanced short-circuit current density up to 14 mA/cm$^2$ and open circuit voltage of up to 650 mV. The device exhibited semi-transparency up to 20% in the wavelength range of 380–740 nm making it suitable for indoor and building-integrated photovoltaic applications. Analytical techniques confirmed the structural and chemical properties of the $Sb_2S_3$ films, demonstrating improved crystallinity and composition with dominant $Sb_2S_3$ contribution above 90 at%. The results underscore the potential developments in magnetron-sputtered $Sb_2S_3$ for emerging transparent thin-film photovoltaics while highlighting the importance of thickness control and interface engineering for efficiency improvements.

# 1. Introduction

Antimony chalcogenides ($Sb_2X_3$, where X = S, Se, or $S_xSe_{1-x}$), as an earth-abundant light-absorbing material class with long-term stability for thin-film photovoltaics has recently been recognized as promising for emerging inorganic thin-film solar cells[1]. $Sb_2X_3$-based compounds have garnered significant research interest within the past decade, with to date the highest reported power conversion efficiencies (PCEs) of 8.0% for $Sb_2S_3$[2], 9.2% for $Sb_2Se_3$[3], and 10.5% for $Sb_2(S_x,Se_{1-x})_3$[4] planar-structured solar cells, all fabricated via solution-based methods. Among $Sb_2X_3$ family, antimony sulfide ($Sb_2S_3$), is one of the most attractive materials for thin-film solar cells due to its favorable optoelectronic properties, including a high absorption coefficient ($\alpha > 10^4$ cm$^{-1}$) and a direct bandgap of around 1.7 eV, while being binary compound with a single stable phase and earth-abundant elemental composition[5]. However, to further develop $Sb_2S_3$-based solar cells, which are yet less explored among $Sb_2X_3$, it requires precise control over film deposition parameters, morphology, crystallinity, and defect passivation[6] to optimize charge transport and reduce recombination losses[7].

Among various fabrication techniques, magnetron sputtering followed by post-sulfurization has emerged as an effective non-solution-based approach for producing $Sb_2S_3$ thin films[8,9]. Sputtering enables precise control over the stoichiometry and thickness of the precursor film, while post-sulfurization enhances crystallinity and phase purity, addressing common issues such as amorphous phases and secondary impurity formation. In this work, we present the first superstrate $Sb_2S_3$ thin-film solar cell fabricated by RF magnetron sputtering with a comparable PCE of 2.76%, demonstrating the viability of this deposition approach for device fabrication. In this study, we report on the fabrication of superstrate-structured $Sb_2S_3$ thin-film solar cells using a two-step process (Figure 1, a): i) RF magnetron sputtering of Sb, ii) followed by a post-sulfurization treatment and finalization of the thin-film solar cell structure. There are relatively few studies reporting on the fabrication of efficient devices using RF magnetron sputtering (Figure 1, b).

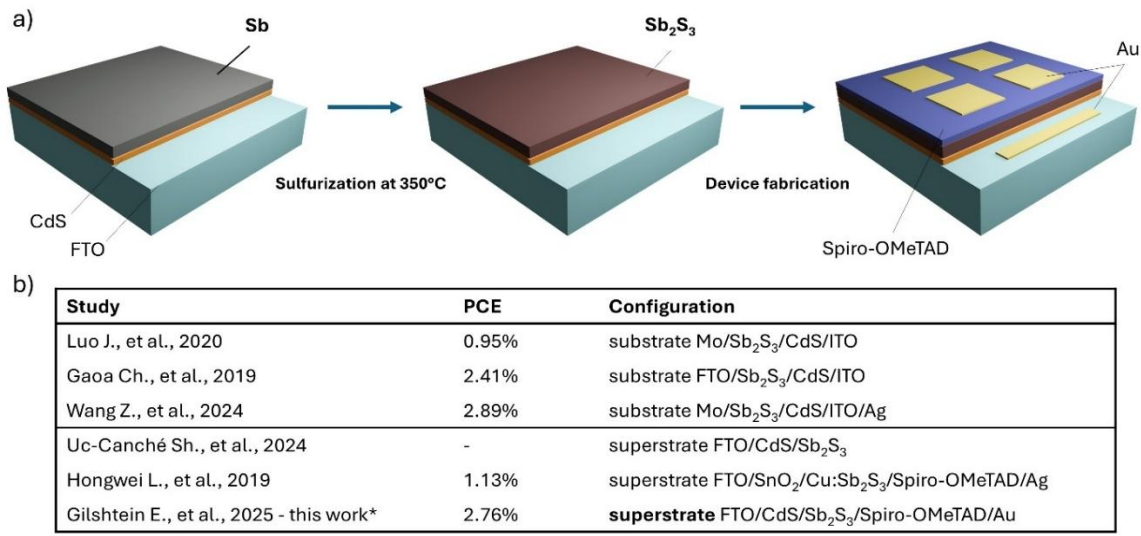

**Fig. 1.** a) Schematics of the solar cell fabrication process used in this study: RF magnetron sputtering of Sb films on glass/FTO substrate with deposited CdS electron transport layer (ETL)/buffer layer, followed by sulfurization of Sb and deposition of Spiro-OMeTAD hole transport layer (HTL) and evaporation of Au contacts, b) Table with reported results on $Sb_2S_3$ thin-film solar cells by magnetron sputtering.

In earlier reported works, substrate-configured $Sb_2S_3$ solar cells obtained by sputtering from high-purity $Sb_2S_3$ targets reached a PCE of 0.95% for $Mo/Sb_2S_3/CdS/ITO$[10], and 2.41% for $FTO/Sb_2S_3/CdS/ITO$[11]. Recently Wang et al.,[12] proved $Mo/Sb_2S_3/CdS/ITO/Ag$ planar substrate-type solar cells with maximum PCE of 2.89% with absorbers made via sputtering from the Sb precursors and post-sulfurization at high-temperature in graphite box. In contrast, a few studies have investigated superstrate-structured devices based on sputtered $Sb_2S_3$. While Hongwei et al. demonstrated solar cells with Cu-doped $Sb_2S_3$ absorbers with 1.13% PCE[13], another study on $FTO/CdS/Sb_2S_3$ configuration reported no efficiency[14]. In this study, we are reporting the superstrate-structured $Sb_2S_3$ solar cells obtained by sputtering with the optimized thicknesses of the buffer layer (CdS) and the absorber layer ($Sb_2S_3$), which is demonstrated to directly impact the short-circuit current density ($J_{sc}$) and open circuit voltage ($V_{oc}$) of the devices. Through systematic investigation, we identified an optimal $Sb_2S_3$ absorber thickness of 100 nm, which not only enhances device performance but also results in semi-transparent thin-film solar cells, highlighting the potential of sputtered $Sb_2S_3$ for applications in photovoltaics and building-integrated photovoltaics (BIPV).

## 2. Experimental section
### 2.1. Preparation of $Sb_2S_3$ thin-film solar cells

FTO-coated glass substrates (Sigma-Aldrich, ~7 Ω/sq) were cleaned using the following protocol: ultrasonic bath for 10 min at 60°C in DI water and 10 min in isopropanol, followed by UV-cleaning for 10 min. Next, the part of the FTO surface (5 mm in width) was covered by the Kapton tape for the Au deposition at the later stage, and the CdS buffer layer was deposited onto the exposed FTO-coated glass substrates using the chemical bath deposition method via a procedure described previously[15]. In brief, the deposition solution consisted of deionized water, 5.98 mM of cadmium sulfate ($CdSO_4$), 29.98 mM of thiourea ($CH_4N_2S$), 19.98 mM of ammonium chloride ($NH_4Cl$), 25 mL of ammonium hydroxide ($NH_4OH$) at a concentration of 28-30%. The samples were placed in 300 mL of solution, heated to 70°C, and stirred continuously for 7 min (to result in 100 nm CdS film) and 4.5 min (50 nm CdS film)[15]. After deposition, the samples were rinsed with de-ionized water and dried using nitrogen gas, followed by the hot plate annealing in the air for 10 min at 400°C. Afterward, Sb metallic films were deposited onto the dried substrates via RF magnetron sputtering from the Sb target (Stanford Advanced Materials, 2"). The working pressure of the sputtering chamber was $5.2 \cdot 10^{-3}$ mbar, and the argon was introduced at a flow rate of 40 sccm. The sputtering power was maintained at 35 W during the whole deposition process, for the sputtering durations set as 10 min, 6.5 min, and 3.3 min to study the effect of precursor thickness on the device performance (so-called "300 nm", "200 nm" and "100 nm" $Sb_2S_3$). The deposition rate was estimated to be ~30 nm/min, with the continuous substrate rotation with 0.2 rad/s for uniform deposition.

Subsequently, the Sb films and 0.1 g of high-purity (> 99.95%) sulfur powder were placed together into a graphite box loaded inside the middle section of a quartz tube furnace for sulfurization. The value of 0.1 g for sulfur powder was selected based on earlier reported work by Wang Z et al.,[12]. We tried the increased the amount of sulfur and tested 0.3 g during sulfurization. However, the performance of solar cells was worse than when using 0.1 g S (Figure S1). The tube was pumped to $2·10^{-3}$ mbar before introducing high-purity $N_2$ gas, with the pressure set inside the tube for 175 mbar during the annealing process (reaching 230 mbar at the peak annealing temperature). The temperature of the whole tube furnace with graphite box containing samples and sulfur powder was increased with 20°C/min ramp rate and kept at 350°C for 15 min with natural cool down by the end of the process. The temperature of 350°C was selected based on previous knowledge[16], and after testing 300°C and 400°C annealing temperatures for our sputtered Sb sulfurization (Figure S2). After the preparation of the crystalline $Sb_2S_3$ absorber layer, the deposition of Spiro-OMeTAD HTL was carried out using the spin-coating method. The solution, which was prepared by mixing 36.6 mg/mL of Spiro-OMeTAD powder in chlorobenzene with 14.5 μL of t-BP, and 9.53 μL of Li-TFSI salt (520 mg/mL in acetonitrile), was stirred for 3 h and spin-coated at 3000 rpm for 30 s. The samples were then dried in the air at 110°C for 10 min on the hot plate. After that, the 95 nm Au electrodes were simultaneously deposited on both the protected side of the FTO substrate for front contact and on the Spiro-OMeTAD layer for back contact via thermal evaporation. Thus, a superstrate configuration of glass/FTO/CdS/$Sb_2S_3$/Spiro-OMeTAD/Au was fabricated for the $Sb_2S_3$-based solar cells with the active area of each solar cell device of 9 $mm^2$.

## 2.2. Characterization

The morphology of the films and cross-sectional images of the solar cell structure were determined using the scanning electron microscope (SEM) Zeiss Merlin under a 5 kV accelerating voltage. The Raman spectroscopy measurements were performed on the Sb precursor films and $Sb_2S_3$ absorber layers with an excitation wavelength of 532 nm and a power of 1 mW using a modified Renishaw Raman spectrometer equipped with a Si CCD detector in confocal mode using a 40X objective lens. The laser spot size was 0.9 μm with an exposure time of 7 s and 10 frames, resulting in 10 acquisitions per spectrum. The X-ray diffraction patterns were performed with an Automated Multipurpose X-ray Diffractometer XRDynamic 500 by Anton Paar equipped with a Primux 3000 sealed-tube Cu X-ray source ($K_{\alpha 1}$=1.5406 Å and $K_{\alpha 2}$=1.5444 Å) and a Pixos 2000 solid-state hybrid pixel detector. The XRD patterns of the samples were acquired using a Bragg−Brentano geometry in continuous mode with a scan speed of 0.02° s$^{-1}$ in a 2θ range from 10° to 60°. The current density-voltage (J-V) characteristic curves of the solar cells were measured at near-standard test conditions (STC: 1000 W/m$^2$, AM1.5, and 25°C), with the use of Newport class ABA steady-state solar simulator. The irradiance was measured with a 2x2 cm$^2$ mono-Si reference cell from ReRa certified at STC by the Nijmegen PV measurement Facility. The J-V measurements were performed using a 4-probe method using a 2400 Keithley source meter, with the voltage swept from -0.2 V to 0.9 V. The external quantum efficiency (EQE) spectra were obtained using the QEX10 Measurement System (from PV Measurements Inc.) equipped with a Xenon arc monochromator lamp source set to 65W power. X-ray photoelectron spectroscopy (XPS) of 100 nm best-performing $Sb_2S_3$ absorbers was performed using a Thermo Fisher Nexsa XPS system with an Al $K_\alpha$ 1486.6 eV excitation source, and a flood gun for charge compensation. The energy scale of the XPS was calibrated using the C 1s peak of the surface carbon at 284.8 eV, which is present on the surface of all the samples. The high-resolution XPS scans of Sb 3d and S 2p were collected from the $Sb_2S_3$ films after 200 s of 2 keV, followed by 200 s of 4 keV Ar cluster etching with a cluster size of 2000 Ar atoms. The data were analyzed using CasaXPS software, with a Shirley-type background applied for the S 2p peak deconvolution, and a linear background applied for the Sb 3d peak deconvolution.

## 3. Results and discussion

In this study, we focused on the superstrate glass/FTO/CdS/$Sb_2S_3$/Spiro-OMeTAD/Au solar cell structure. Figure 2 illustrates the optimization of the CdS ETL thickness, as it plays a critical role in achieving improved device performance. Initially, the absorber layer thickness was fixed to 300 nm because the thicker $Sb_2S_3$ films (300 nm – 1 μm) have defective double-layer films (Figure S3) and yielded non-functional solar cells. Insufficient light absorption is known to occur if a too-thin $Sb_2S_3$ layer is used, while an over-thick $Sb_2S_3$ layer may cause the carriers recombination by the defects over the large transportation distance, becoming a more serious drawback than insufficient light absorption[17]. In Figure 2a, the schematic illustrates two device configurations based on a 300 nm $Sb_2S_3$ absorber: one with a 100 nm CdS ETL and another with a reduced 50 nm CdS on the FTO-coated glass.

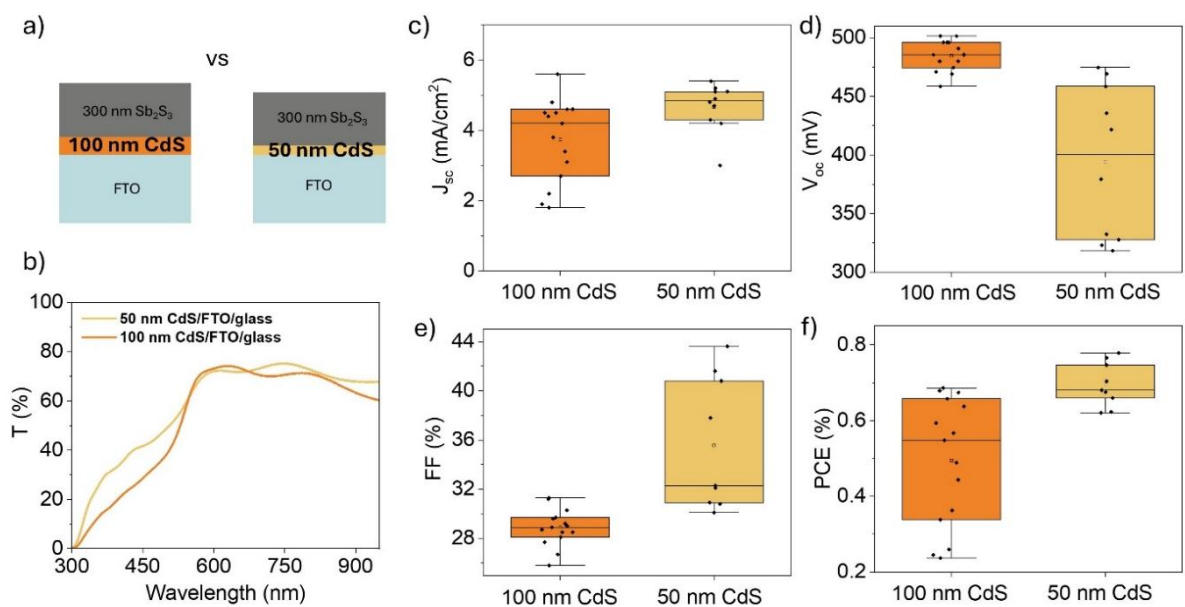

**Fig. 2. Effect of CdS layer thickness on device performance:** a) Schematics of the two device configurations with 100 nm and 50 nm CdS layer; b) Transmittance curves of glass/FTO/CdS with 100 nm and 50 nm layers stack; c) Short circuit current density ($J_{sc}$), d) Open circuit voltage ($V_{oc}$), e) Fill factor (FF), and f) Power conversion efficiency (PCE) of the two superstrate device types glass/FTO/CdS/Sb$_2$S$_3$/Spiro-OMeTAD/Au with 100 nm and 50 nm CdS layer.

The transmittance spectra of the glass/FTO/CdS (measured over the air as reference) with 100 and 50 nm CdS films in the wavelength range from 300 to 950 nm are shown in Figure 2b. The oscillations beyond the absorption edge are multiple reflections at CdS/FTO, FTO/glass, and CdS/air interfaces[18]. The spectra show an average T% of over 70% beyond the absorption edge for both CdS layer thicknesses, with a major part of the visible light passing the glass/FTO/CdS layers stack. However, the 100 nm thick CdS layer is less transparent in the UV part of the spectra, which results in increased light absorption by the glass/FTO/CdS before it reaches the Sb$_2$S$_3$ absorber. Using both CdS layer thicknesses, we fabricated the solar cells by consecutive deposition of 300 nm Sb using RF magnetron sputtering, sulfurization in the tube furnace, spin-coating the Spiro-OMeTAD HTL, and Au evaporation, as described in detail in the Experimental section. Figure 2 (c-f) presents the main photovoltaic parameters ($V_{oc}$, $J_{sc}$, FF, and PCE) of the two devices, revealing an improvement in photovoltaic performance for the device with the thinner 50 nm CdS layer. The reduction in CdS thickness decreases parasitic absorption in the ETL, allowing more light to reach the Sb$_2$S$_3$ absorber, which aligns with T% measurements. The statistical data shows enhancements in $J_{sc}$, FF, and overall efficiency (PCE) for the 50 nm CdS device. The only parameter that has decreased from the average of 475 mV to 400 mV with the decreased CdS layer thickness is the $V_{oc}$. This might be because a thinner

CdS layer is less effective in passivating surface defects at the interface, altering the conduction band offset and, therefore, increasing interface recombination. However, for the next experiments, we fixed the thickness of CdS to 50 nm, as more promising in terms of PCE.

As the next step, we analyzed the influence of the $Sb_2S_3$ absorber thickness on the photovoltaic properties of the solar cells based on absorbers with 300 nm, 200 nm, and 100 nm thicknesses deposited on a 50 nm CdS ETL on FTO-coated glass (Figure 3 a). For simplicity, the samples were titled "300 nm", "200 nm", and "100 nm" while the actually measured thicknesses were 350 ± 20 nm, 230 ± 20 nm, and 140 ± 20 nm, respectively.

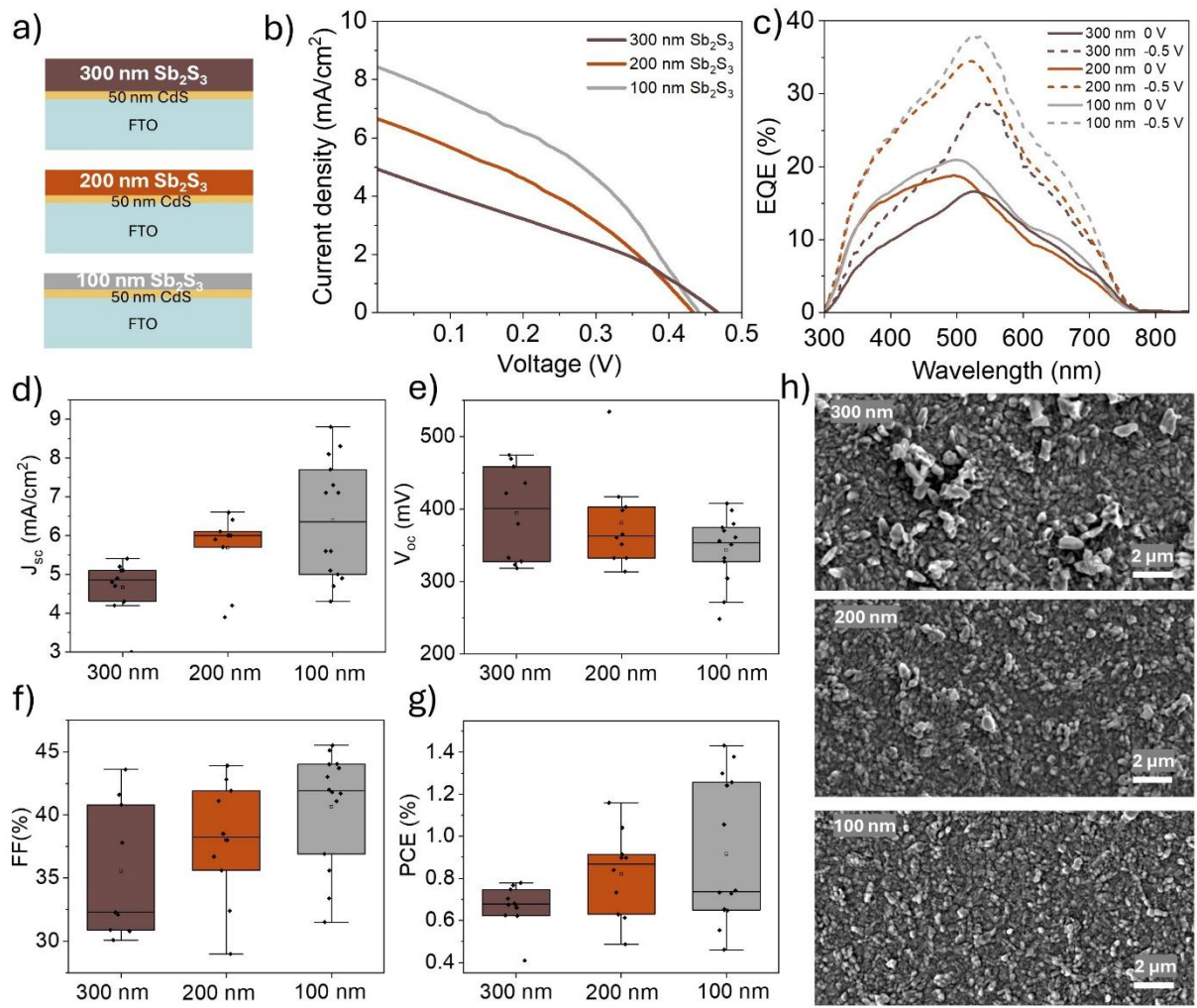

**Fig. 3. The impact of the $Sb_2S_3$ absorber thickness on photovoltaic performance of solar cells:** a) Schematics of the three device configurations with 300, 200, and 100 nm $Sb_2S_3$ (all with 50 nm CdS); b) JV curves and c) EQE spectra for glass/FTO/CdS/$Sb_2S_3$/Spiro-OMeTAD/Au solar cells with different absorber thicknesses. Statistical analysis of d) $J_{sc}$, e) $V_{oc}$, f) FF, and g) PCE of the devices with 300, 200, and 100 nm $Sb_2S_3$; h) SEM top-view images of the 300, 200, and 100 nm $Sb_2S_3$ films morphology.

The J-V curves (Figure 3 b) show an increase in device performance metrics, particularly the $J_{sc}$ (from 4.9 to 8.4 mA/cm$^2$), with a decrease in the absorber thickness from 300 to 200 and 100 nm, while the $V_{oc}$ remains almost unchanged (432 - 464 mV). It has been demonstrated earlier that increasing the absorber layer thickness to a certain point enhances the photocurrent[17]. However, when the thickness exceeds the hole diffusion length (180 nm in $Sb_2S_3$[19]), the inferior area in $Sb_2S_3$ for an inefficient hole generation and transport decreases the $J_{sc}$ and external quantum efficiency (EQE). This reduction is also evident in the EQE spectra (Figure 3 c), where the thinner absorbers exhibit reduced light harvesting. The peak EQE value reached over 20% for solar cells with 100 nm absorbers. On the other hand, the peak value of the EQE spectrum for a 200 nm $Sb_2S_3$ device reached 18%, and for a 300 nm $Sb_2S_3$ device could hardly reach 15%. This is unlikely to be the case for the increased absorber thickness, typically resulting in the development of the proper stoichiometry and fewer defect states in the material[20]. In the case of our sputtered and post-sulfurized $Sb_2S_3$ absorbers, there is a noticeable improvement in the charge extraction and reduction of charge-trapping effects for thinner absorbers (100 nm). Thicker $Sb_2S_3$ 300 nm layers may suffer from higher recombination losses because photo-generated carriers in the 300–500 nm range are created closer to the CdS/$Sb_2S_3$ interface, where defects and recombination rates are higher[21]. The carriers must travel farther to reach the Au contact, increasing the probability of recombination. Also, thinner absorbers allow better transmission of short-wavelength photons into the active layer, leading to more efficient light utilization in this spectral range. Long-wavelength photons penetrate deeper into the absorber material, so the photo-generated carriers in the 500–800 nm range are less affected by surface recombination at the CdS/$Sb_2S_3$ interface. The overall EQE in this range depends more on bulk properties, which are most likely similar for 100, 200, and 300 nm absorbers. We also calculated the series resistance ($R_s$) and shunt resistance ($R_{sh}$) of the devices[22] with different $Sb_2S_3$ absorber thicknesses. The $R_s$ values are 52 Ω·cm$^2$, 35 Ω·cm$^2$, and 25 Ω·cm$^2$ for 300, 200, and 100 nm absorbers, respectively, while $R_{sh}$ values are 114 Ω·cm$^2$, 110 Ω·cm$^2$, and 102 Ω·cm$^2$ for 300, 200, and 100 nm absorbers. Both $R_{sh}$ and $R_s$ decrease with the absorber thickness reduction. A decrease in $R_s$ with thickness decrease has a positive aspect minimizing resistive losses meaning better charge transport within the device due to the reduced distance charges needed to travel. Lowering the $R_{sh}$ with thickness decrease has a negative aspect meaning more recombination pathways at interfaces, reducing the $V_{oc}$. At the same time, the values of $R_{sh}$ for all the devices are still far away from the desirable values in kΩ·cm$^2$ range[23], therefore the change from 114 to 102 Ω·cm$^2$ for 300 nm to 100 nm absorbers is almost negligible in this range. The main photovoltaic properties such as average $J_{sc}$, FF, and PCE (Figure 3 d, f, g) are found to be improved for the devices with the decreased absorber thickness (200 and 100 nm $Sb_2S_3$). In contrast, the $V_{oc}$ (Figure 3 e) is slightly decreased from 400 to 350 mV due to the reasons mentioned above. The average values for the main photovoltaics parameters for the $Sb_2S_3$ solar cells with different absorber thicknesses are summarized in Table 1. For the 100 nm absorber, it was possible to achieve above 1.4% PCE and above 40% FF for these thickness series.

|  | $J_{sc}$ (mA/cm$^2$) | $V_{oc}$ (mV) | FF (%) | PCE (%) | $R_s$ (Ω·cm$^2$) | $R_{sh}$ (Ω·cm$^2$) |
|---|---|---|---|---|---|---|
| 300 nm $Sb_2S_3$ | 4.7 | 393 | 36 | 0.67 | 52 | 114 |
| 200 nm $Sb_2S_3$ | 5.7 | 380 | 38 | 0.82 | 35 | 110 |
| 100 nm $Sb_2S_3$ | 6.4 | 343 | 41 | 0.92 | 25 | 102 |

**Table 1**. The average photovoltaic metrics for thin-film solar cells based on 300, 200, and 100 nm $Sb_2S_3$ absorbers.

It is known that both the quality and thickness of the absorber layer are very important factors that will affect the absorption of incident light, carrier transport, and power conversion efficiency in solar cells. Correspondingly, the SEM images (Figure 3, h) reveal structural differences, with thinner $Sb_2S_3$ layers (100 nm) showing a smoother surface morphology compared to the thicker layers (200 and 300 nm), which exhibit non-uniform agglomerations on top of the film, unfavorable for the formation of smooth interface with the Spiro-OMeTAD layer. For the 100 nm $Sb_2S_3$ film, the average grain size is 220 nm. Larger grain size and fewer grain boundaries would be more beneficial for the $Sb_2S_3$ thin films, with reported grain lateral size values of 1–2 μm[24] providing better compactness of thin films. Nevertheless, for the tested configuration, the results collectively suggest reducing $Sb_2S_3$ thickness to 100 nm allows better carrier collection, improving overall device efficiency.

We further analyzed the composition and structural properties of the 100 nm $Sb_2S_3$ layer synthesized via Sb sputtering followed by post-sulfurization. High-resolution XPS spectra of the Sb 3d and S 2p core levels are presented in Figure 4b and c, respectively. After 400 s of Ar$^+$ cluster etching, the Sb 3d$_{5/2}$ spectrum revealed the presence of antimony in three distinct chemical states: $Sb_2S_3$ (530.3 eV), metallic Sb (529.7 eV), and $Sb_2O_3$ (531.1 eV)[25]. The corresponding Sb 3d$_{3/2}$ peaks were also observed, with a characteristic doublet separation of 9.34 eV. Additionally, an O 1s peak was detected at 532.0 eV. Quantitative analysis indicates that Sb-S bonds dominate the composition of the layer, accounting for approximately 79 at%, while Sb-Sb and Sb-O species contribute 12 at% and 9 at%, respectively. The presence of metallic Sb suggests that the sample may contain unreacted or segregated antimony from the Sb precursor film. Along with it, Ar$^+$ cluster etching can induce reduction reactions in the material, breaking Sb-S bonds and leading to the formation of metallic Sb[26]. The presence of $Sb_2O_3$ can result from oxidation during the air exposure of the samples, along with the effect of Ar$^+$ etching causing chemical changes by inducing oxidation due to residual oxygen in the chamber[27]. Sulfur can be detected in two states: one doublet with binding energies at 162.9 and 161.8 eV - characteristic of sulfur in the metal sulfides ($Sb_2S_3$), which has a dominant contribution above 90 at% in the layer, and another S state at 164.7 and 163.6 eV (S-S species)[25]. The minor S–S peak (below 10 at%) indicates the presence of elemental sulfur ($S_8$ clusters)[28] from unreacted or segregated sulfur in the layer from the sulfurization process.

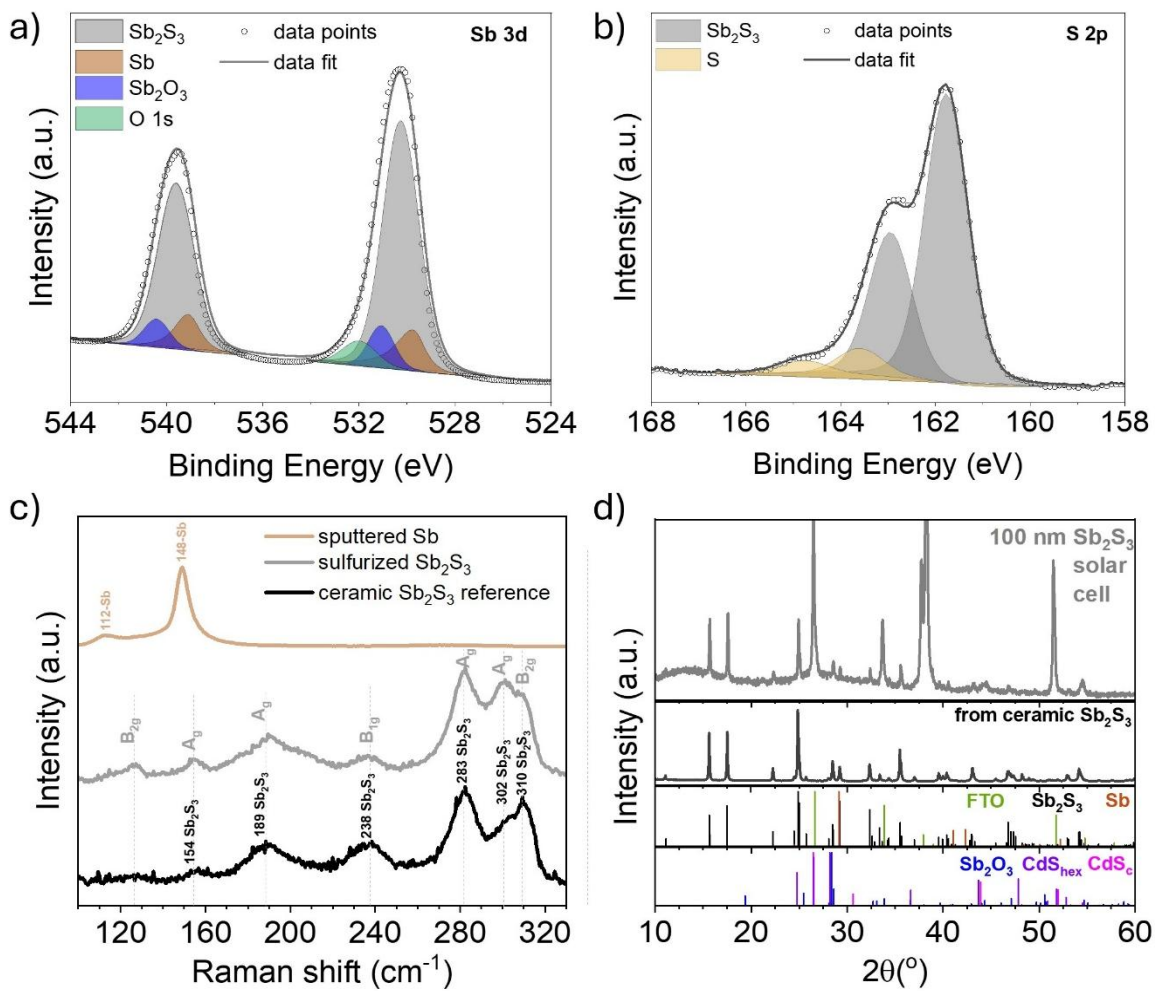

**Fig. 4. XPS, Raman, and XRD characterization of 100 nm $Sb_2S_3$ absorber:** XPS high-resolution spectra of a) Sb 3d and b) S 2p after 400 s Ar$^+$ ion sputtering; c) Raman spectra of the Sb metallic film and $Sb_2S_3$ film sulfurized from Sb, vs. $Sb_2S_3$ from ceramic target reference measured with 532 nm laser excitation wavelength; d) XRD patterns of fabricated $Sb_2S_3$ film and reference reflexes according to annotations of stibnite $Sb_2S_3$, FTO, CdS, $Sb_2O_3$, and compared with ceramic $Sb_2S_3$ target.

The six Raman lines are assigned to the $Sb_2S_3$ phase as in Fig. 4c, namely at 154 ($A_g$), 189 ($A_g$), 238 ($B_{1g/3g}$), 283 ($A_g$), 302 ($A_g$), and 310 cm$^{-1}$ ($B_{2g}$) which further confirms the $Sb_2S_3$ crystallization[29–31] when compared to the XRD pattern from the ceramic $Sb_2S_3$ reference target (grey line). Before sulfurization, $E_g$ (112 cm$^{-1}$) and $A_{1g}$ (148 cm$^{-1}$) modes of antimony vibrations[33] are present in the sputtered Sb film (light-brown line). Along with 532 nm, in this study we also used another excitation laser wavelength (633 nm) for Raman measurements. It allowed higher Raman intensity, being closer to resonance with electronic transitions in $Sb_2S_3$ and making it a more effective probe for Raman analysis of these films with higher signal-to-noise ratio (Figure S4)[34]. Peak-specific resonance enhancement is clearly observed with 633 nm excitation, especially for $A_g$ modes. Several characteristic $Sb_2S_3$ peaks (e.g., at 154 cm$^{-1}$, 189 cm$^{-1}$, 237 cm$^{-1}$, and 282 cm$^{-1}$) appear stronger with 633 nm excitation due to the above-mentioned resonance Raman effect ($Sb_2S_3$ bandgap corresponds more closely to the 633 nm – 1.96 eV photon energy). The 114 cm$^{-1}$ characteristic peak of $Sb_2O_3$ is also enhanced by the resonance condition at 633 nm, making it slightly visible only under this excitation (Figure S4). This also might be caused by longer wavelengths deeper penetrating into the sample, which allowed to probe minor surface or sub-surface oxidation of $Sb/Sb_2S_3$ during the synthesis or storage process. This was also confirmed by the XPS described earlier, with Sb-O species contributing with 9 at% after $Ar^+$ sputtering for surface cleaning. Thereby, different modes couple differently with different excitation energies, providing complementary information.

The crystallinity and orientation of the 100 nm $Sb_2S_3$ thin film on CdS/FTO-coated glass are studied using X-ray diffraction (XRD) (Figure 4 d). The XRD pattern of the $Sb_2S_3$ film was compared with the polycrystalline anisotropic oriented ceramic target of $Sb_2S_3$, as well as with standard pattern as per ICDD PDF no. 00-042-1393 (space group: *Pbnm*; $a$ = 11.2390 Å; $b$ = 11.3130 Å; $c$ = 3.8411 Å and $\alpha = \beta = \gamma = 90°$)[35]. The strong peak intensities and narrow peak FHWM suggest that the films of $Sb_2S_3$ obtained using the method proposed in this study have a single phase with high crystallinity[36]. From the XRD pattern of the $Sb_2S_3$ on CdS/FTO, it is observed that the systematic absence of *(hkl)* reflexes, *where l ≠ 0*, as well as the relative higher intensity of *(hk0)*-attributed peaks suggest that the majority of the crystallites have their [00*l*] direction perpendicular to the normal to the surface and are dispersed in the film surface plane. The zoomed-in XRD pattern regions (from 14 to 56°) are also shown on Figure S5.

Less critical at first glance parameters, such as the "age" of the CdS "mother solution" (chemicals dissolved in DI water), the cleanliness of the tube furnace in which sulfurization of Sb layers is occurring, the uniformity of the Spiro-OMeTAD layer by a spin-coating process, are in the end significantly influencing all together the quality of the $Sb_2S_3$-based thin-film solar cells. By preparing the fresh (same day) CdS mother solution, cleaning the tube furnace used for sulfurization at 900°C for 30 min to remove residues of S, double filtering the solution of Spiro-OMeTAD HTL using 0.45 μm filter, and again using 0.1 μm PTFE syringe filter we repeated the experiments with 100 nm $Sb_2S_3$ absorbers and fabricated the solar cells according to the same protocol. The results were obtained by repeating a series of samples under identical processing conditions, so that the average $J_{sc}$ of the devices reached 11.4 mA/cm$^2$ (maximum value 14.0 mA/cm$^2$), $V_{oc}$ – 626 mV (maximum of 672 mV), FF – 29% (maximum of 31%) and PCE – 2.06% (maximum value 2.76%) (Figure 5 a).

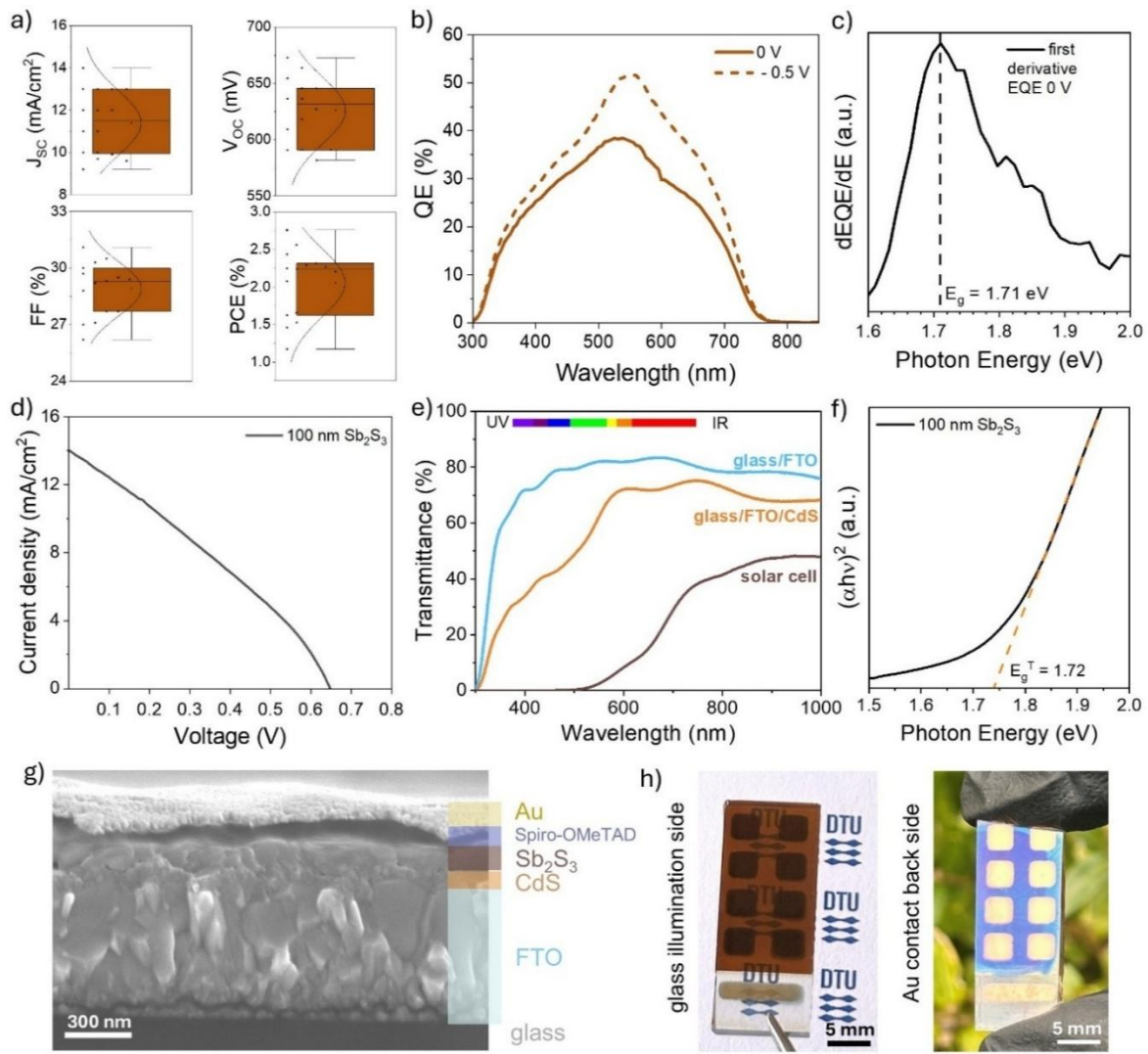

**Fig. 5. The best-performing superstrate structured glass/FTO/CdS/$Sb_2S_3$/Spiro-OMeTAD/Au solar cell devices with 100 nm $Sb_2S_3$:** a) Statistical analysis of $J_{sc}$, $V_{oc}$, FF, and PCE of various cells, b) EQE spectra with no bias and -0.5V applied bias plotted versus the wavelength of the incident light, and c) the derivative of the EQE for optical bandgap $E_g$ extraction. d) J-V curve of the device; e) UV–vis transmittance spectrum of the solar cell device (area between Au contacts), and f) Tauc plot of the optical absorption. g) SEM cross-sectional image of the complete device stack and h) photos of the devices from the back and front sides.

The EQE of the devices exhibited enhanced charge extraction efficiency at the applied negative bias (- 0.5V) helping to reduce recombination losses at the interfaces (Figure 5 b). The maximum of the EQE spectra is shifted from 532 nm (no bias) to 551 nm (-0.5 V) due to the change in the cell's internal electric field distribution, also suggesting that the applied bias enhances sufficiently carrier collection efficiency in more defective deeper regions/interfaces of the cell (toward $Sb_2S_3$/Spiro-OMeTAD/Au interfaces). The peak EQE intensity value significantly improved from 15% to above 35% compared to the previous device series. The first derivative d(EQE)/dE (Figure 5c) exhibits a clear inflection point corresponding to a band gap ($E_g$) of 1.71 eV, which correlates with the values obtained for $Sb_2S_3$[5]. The value of $R_s$ for the 100 nm $Sb_2S_3$ solar cells was significantly improved and lowered to 17 Ω·cm$^2$, while the $R_{sh}$ values remained undesirably low (66 Ω·cm$^2$) as seen from the steep slope of the J-V curve at 0V (Figure 5 d). The average visible transparency of almost 20% was obtained in the wavelength range of 380–740 nm for the glass/FTO/CdS/$Sb_2S_3$ solar cell device stack (Figure 5 e). Based on the optical measurements, the $E_g$ derived from the Tauc plot (Figure 5 f) of 1.72 eV is in good agreement with the one extracted from the EQE. We confirmed via cross-sectional SEM that the solar cells based on 100 nm $Sb_2S_3$ absorbers are free from double layers and pinholes, with smooth interfaces between each layer (Figure 5 g). The thicknesses of the layers were according to the control values mentioned in the Materials and Methods section. As the total thickness of the device stack (FTO/CdS/$Sb_2S_3$/Spiro-OMeTAD) on glass is below 1 μm, and the absorber layer itself is ultrathin (100 nm), the fabricated devices exhibit semi-transparency[37] as shown in the photograph (Figure 5 h), when the glass side is facing up. The dark squares are the Au contact pads on Spiro-OMeTAD at the backside. This feature can find advantages when the solar cells need to be integrated into transparent or translucent surfaces for in-door PV applications. They can harvest ambient light while allowing visible light to pass through, enabling integration into windows, displays, or building surfaces without obstructing natural lighting.

**Conclusion**

In this work, the superstrate-structured $Sb_2S_3$ thin-film solar cells are fabricated via RF magnetron sputtering of thin Sb films followed by post-sulfurization in sulfur-rich tube furnace environment. The two-step fabrication process enabled the formation of crystalline stibnite-phase $Sb_2S_3$ absorbers with controlled thickness. Through systematic optimization, the study demonstrated that absorber thickness is a critical determinant of device performance, with a 100 nm $Sb_2S_3$ layer achieving the best PCE of 2.76%. This enhancement is attributed to improved carrier extraction and reduced recombination losses, as thicker layers (e.g., 300 nm) were found to introduce increased charge carrier recombination due to defects and transport limitations. Similarly, the reduction in the CdS ETL thickness to 50 nm enhanced light transmission to the $Sb_2S_3$ absorber, improving the $J_{sc}$. Raman and XPS analyses further revealed the dominant presence of Sb-S bonds with minor contributions of Sb-Sb and Sb-O species, while SEM demonstrated uniform and pinhole-free films at optimal thickness. Despite the promising results, the study identifies several challenges that remain to be addressed for further efficiency improvement. Key areas for future exploration include advanced defect passivation strategies, interface engineering to minimize recombination losses, and further optimization of the absorber/buffer layer interface. Additionally, enhancing $R_{sh}$ and further reducing the $R_s$ of the device stack is essential for achieving higher fill factors and overall efficiency. This study establishes a baseline for the development of efficient, semi-transparent $Sb_2S_3$ solar cells and highlights the versatility of magnetron sputtering for thin-film PV.

## Acknowledgments

This work is supported by the M-ERA.NET Grant funded by the Innovation Fund Denmark project (*Lightcell,* project number 2118-00014B), and by Independent Research Fund DFF-Research Project 1 - Inge Lehmann (New, tunable n-type material for emerging thin-film solar cells, project number 1134-00005B71336). NIMP authors acknowledge funding from the Romanian Ministry of Research, Innovation and Digitalization through the Core Programme PC3-PN23080303 project, and from UEFISCDI through ERANET-M-3-ERANET-Ligthcell project (contract No. 19/15.03.2024). The authors would also like to acknowledge the contributions of DTU MSc students Fardin Ghaffari-Tabrizi and Ritjiua Kahuure for preliminary work on Sb and $Sb_2S_3$ sulfurization. We also thank Moises Espindola for training on EQE system and Au evaporation setup, together with Professor Jens Wenzel Andreasen from DTU Energy for kindly allowing us to use the laboratory facilities and measurement equipment.

**Supplementary Figures**

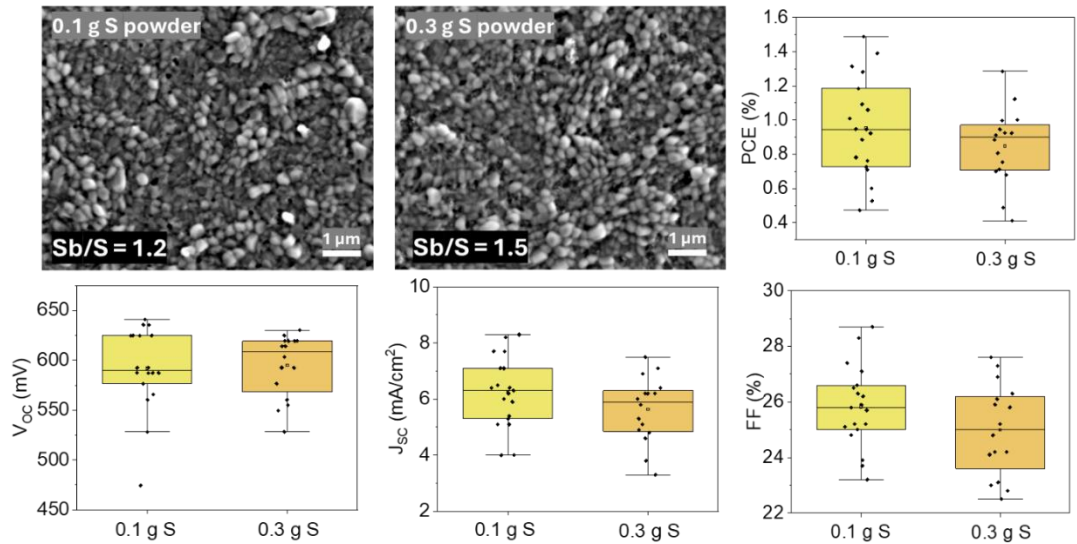

**Fig. S1**. Variation of S powder amount during sulfurization of Sb: SEM images revealed no changes in the morphology of the films (0.1 g vs 0.3 g); the PCE, $V_{oc}$, $J_{sc}$, and FF are lowered when 0.3 g of S powder is used compared to 0.1 g.

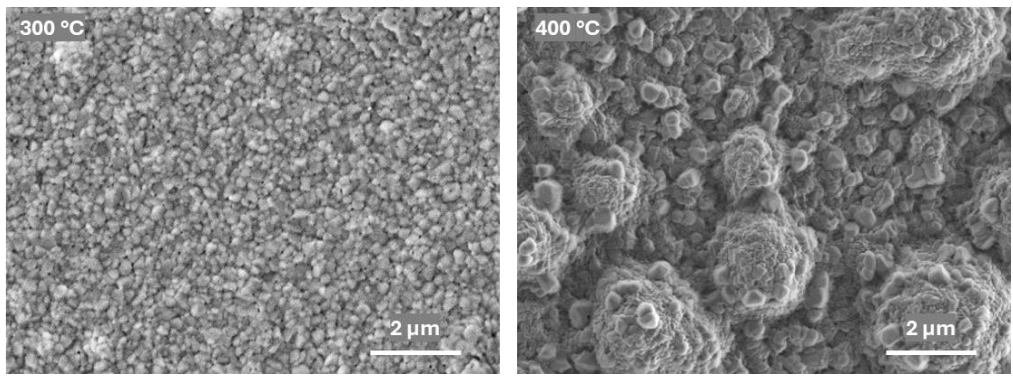

**Fig. S2.** Graphite box annealing temperature variation: SEM images of the $Sb_2S_3$ films after sulfurization at 300 (revealed amorphous film with fewer pronounces grains) and 400°C (revealed inhomogeneous grains and clusters grown perpendicular to the plane direction).

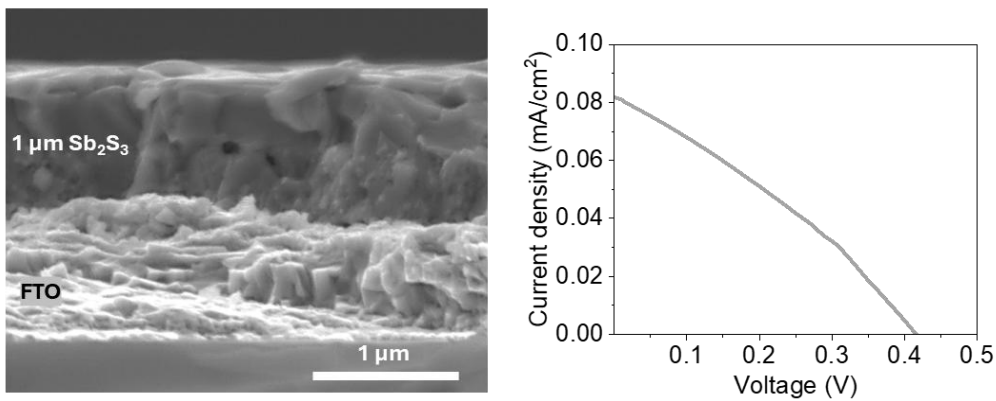

**Fig. S3**. The double-layer formation for the $Sb_2S_3$ 1 μm film thickness, and corresponding J-V curve for the device with < 0.01% PCE based on it.

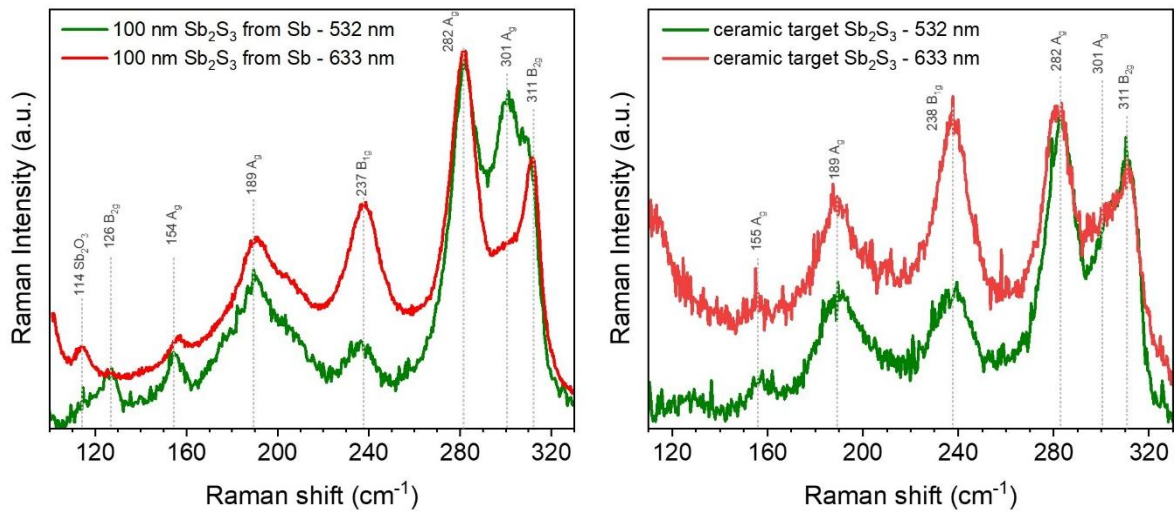

**Fig. S4.** Multiwavelength Raman spectra of the Sb$_2$S$_3$ films obtained by sulfurization of sputtered Sb compared to the films obtained from the ceramic Sb$_2$S$_3$ reference target.

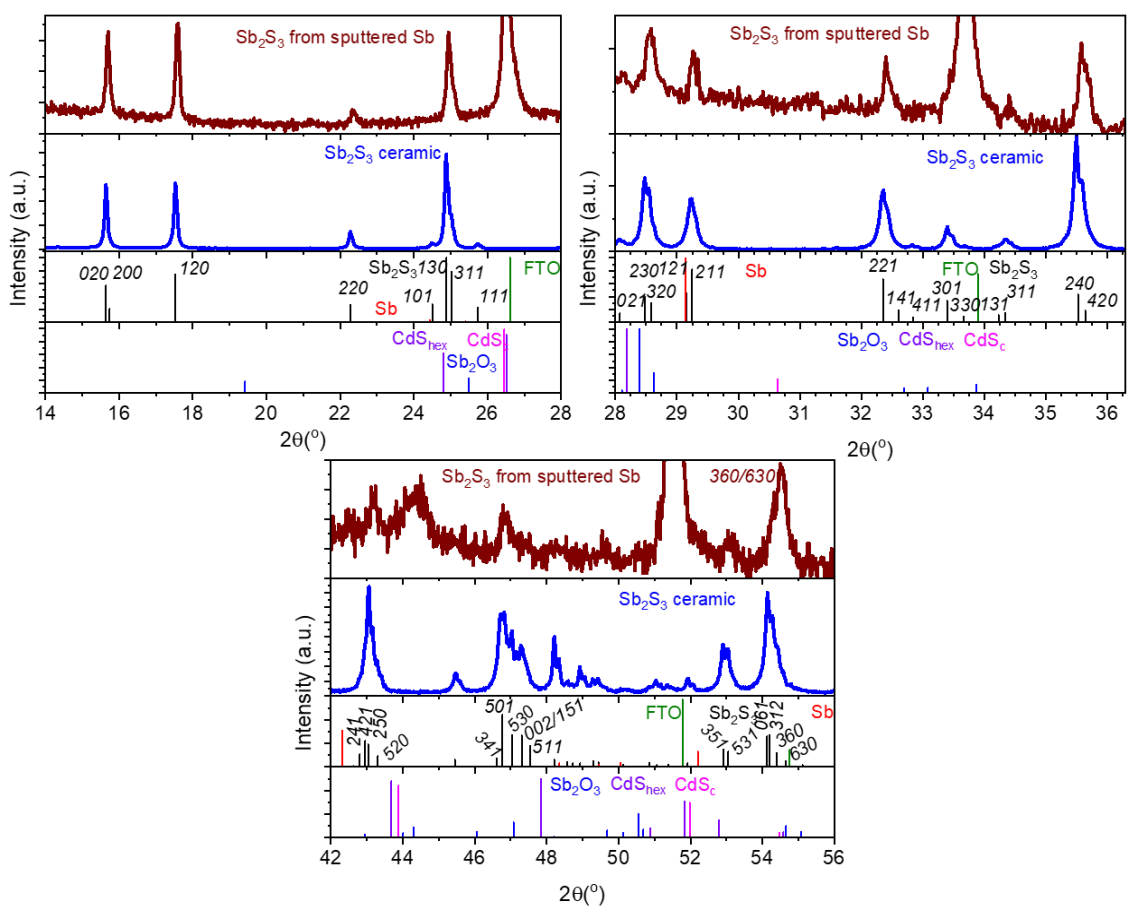

**Fig. S5.** Zoom-in XRD patterns of the Sb$_2$S$_3$ films from sulfurized Sb fabricated in this study in a 2-theta range of 14 to 56° compared to the reference Sb$_2$S$_3$, Sb$_2$O$_3$, Sb, as well as FTO, and CdS reflexes.